# Orbital angular momentum and informational entropy in perturbed vortex beams

A. Volyar, M. Bretsko, Ya. Akimova, Yu. Egorov

The first simple experiments of self-reconstructing or self-healing Bessel vortex beams after being obscured by symmetric obstacles [1,2] caused a flow of publications, for example, on microscopy with self-reconstructing beams [3], the information transmission and data processing systems [4,5]. Self-healing properties were also found in other types of vortex beams [6-8]. However, a more careful analysis of the self-healing showed [9] that Bessel beams do not always self-heal from transparent obstacles, and never self-heal from turbulence. This article caught our attention because it raises the question: what parameters of the beam characterize its self-reconstruction? The answer to this question depends on the use of feature properties of the vortex beam in various optical devices. For example, the sector obstacles inject some measure of uncertainty into the vortex beam [12-14] between the orbital angular momentum (OAM) and the angle of the sector obstacle, which, in turn, triggers a chain of vortex birth and annihilation events that increases a number of new vortex states. But such a perturbation although causing irreversible destruction of the wavefront, allows one to determine the topological charge of the optical vortex [10] and the orbital angular momentum [11]. On the other hand, the information properties of the vortex beam as a whole deteriorate significantly. Indeed, the growth of internal uncertainty in the wave structure induced by the sector aperture iindicates significant changes in such physical characteristics as informational entropy [15] and spatial coherence [16]. The fundamental problems of analyzing informational entropy (or Shannon's entropy) in light beams were considered in detail by Francis [15] and applied to vortex beams by the authors of Ref. [17,18]. The authors of Ref.[17] theoretically estimated the informational entropy and pointed out the relationship between entropy and spatial coherence for the 1D case of an unperturbed quasi-monochromatic vortex beam while the authors of Ref. [18] presented experimental confirmation of these theoretical predictions. We paid attention to the fact that, in the general case, informational entropy characterizes the vortex beam as a whole rather than its 1D projection onto the observation plane, as in Ref.[17, 18]. Besides, the main contribution to information entropy is made by changes in the number of vortex states in the beam subjected to external perturbations. It is such an approach allowed the authors of Ref. [13] to investigate the relationship between the uncertainty of the angular position inside the vortex beam and the OAM. These studies point out the relationship between the vortex spectrum of the perturbed beam, the OAM, and the information entropy. Thus, the purpose of our letter is both a computer simulation and experimental measurement of informational entropy and the OAM of the perturbed vortex beam on the base of analyzing their vortex spectra.

**1.** First, we analyze the vortex spectrum that occurs after passing the Laguerre-Gauss beam through a hard-edged sector diaphragm. The basic idea is that the sector aperture causes a significant distortion of the wavefront along the azimuthal and radial directions. This results in originating a wide range of optical vortices with different OAM. Thus, the main task is to analyze the OAM and information entropy depending on the vortex spectrum of the perturbed beam.

Let us consider passing a scalar Laguerre-Gaussian $LG_0^m$ beam with an azimuthal index (a topological charge) $m$ and a zero radial index $p=0$ through the opaque regular sector obstacle with an angle $\alpha$ shown in Fig. 1. The vertex sector touches the beam axis. The beam field at the initial plane $z=0$ is represented as

$$\Psi_m(r,\varphi,\alpha) = (\rho/w)^{|m|} e^{im\varphi} e^{-\rho^2/w^2} = r^{|m|} e^{im\varphi} e^{-r^2}, \quad \alpha < \varphi < 2\pi - \alpha, \qquad (1)$$

where $r = \rho/w$. $w$ is a beam waist radius at the plane $z=0$, $\rho$ and $\varphi$ are polar coordinates. We write the beam field (1) modulated by the hard-edged aperture with the angle $\alpha$ as a series of non-normalized Laguerre-Gauss beams $LG_p^m$ in the form

$$\Psi_m(r,\varphi,\alpha) = \sum_{n=-\infty}^{\infty} C_{m,n}(\alpha) LG_0^n(r,\varphi) = \sum_{n=-\infty}^{\infty} C_{m,n}(\alpha) r^{|n|} e^{in\varphi} e^{-r^2}, \quad (2)$$

where the beam amplitudes are

$$C_{m,n}(\alpha) = (-1)^{m-n} \Gamma\left(\frac{|m|+|n|}{2}+1\right) \frac{\sin[(m-n)(\pi-\alpha)]}{m-n} / \left(\pi 2^{\frac{|m|-|n|}{2}} |n|!\right), \quad (3)$$

$\Gamma(x)$ is a Gamma function. The terms in the series (2) with radial indices $p \neq 0$ disappear due to the LG mode orthogonality. The perturbed beam (2) far from the initial plane $z=0$ is written as

$$\Psi_m(r,\varphi,z,\alpha) = \sum_{n=-\infty}^{\infty} C_{m,n}(\alpha) \left(\frac{r}{\sigma(z)}\right)^{|n|} e^{in\varphi} e^{-\left(\frac{r}{\sigma}\right)^2} / \sigma(z), \quad (4)$$

where $\sigma(z) = 1 - iz/z_0$, $z_0 = kw^2/2$, $k$ is a wavenumber.

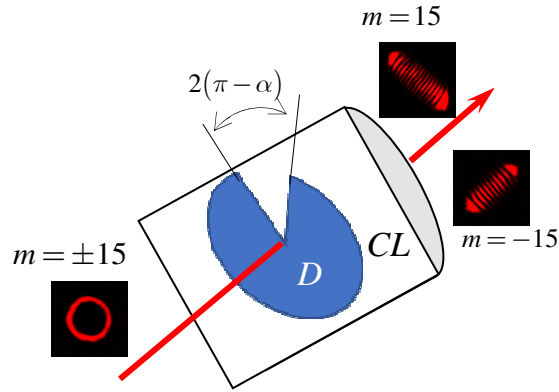

Fig. 1 Sketch of the hard-edged aperture (D) installed at the cylindrical lens plane (CL). Images LG and HG beams illustrate the astigmatic transformation of a single vortex beam.

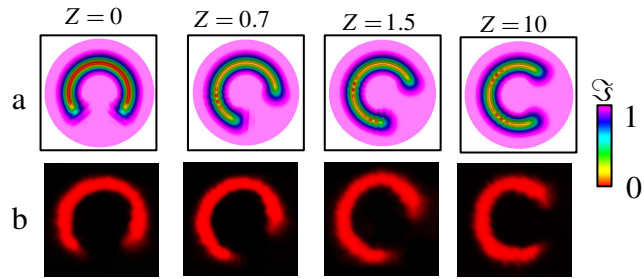

Fig.2 Intensity distribution $\Im(r,\varphi,z)$ of the vortex beam with $m=15$ along length $Z=z/z_0$ perturbed by a hard-edged sector diaphragm with $\alpha = \pi/4$: (a) theory, (b) experiment, the beam waist radius $w_0 = 0.5 mm$, $\lambda = 0.6328 \mu m$

Fig. 2 shows a computer simulation of the intensity distribution $\Im(r,\varphi,z)$ in Eq.(4) along the vortex beam length $(m=15)$ modulated by the sector aperture with $\alpha = \pi/4$. It is noteworthy that the sector aperture cuts out dark sector with blurred edges in the beam intensity distribution that do not heal itself, The dark sector turns synchronically when propagating the beam. We found, both theoretically and experimentally, that beam self-healing does not occur at any beam length $z$ even at a very small sector angle $\alpha$.

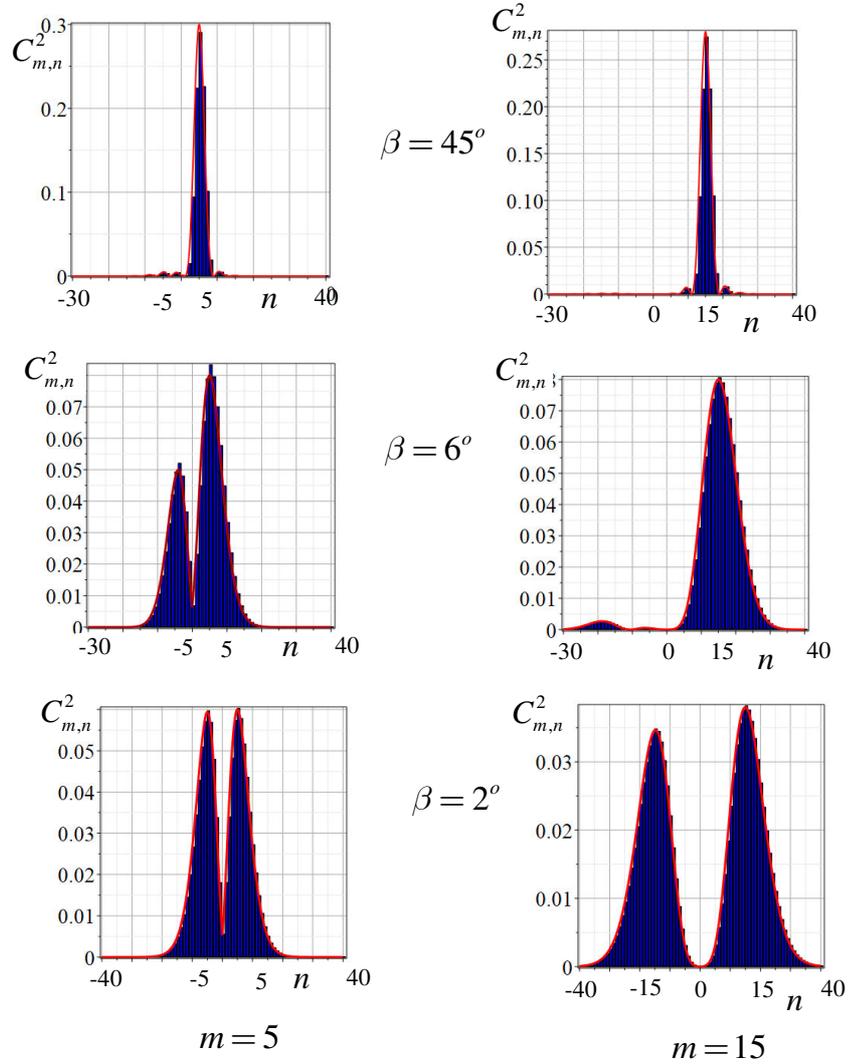

$m=5$  $m=15$

Fig.3 Vortex spectra $C^2_{m,n}$ of the vortex beams with initial topological charges $m=5$ and $m=15$ perturbed by the hard-edged aperture with the sector adjacent angle $\beta = \pi - \alpha$, The solid line is a spectral envelope

*Vortex spectrum.* As we have shown in Ref.[19-21], the analysis of the vortex spectrum (dependence of the vortex mode intensity $\overline{C}^2_{m,n}(\alpha)$ on the topological charge $n$) enable us recovering the initial beam shape and finding its OAM. In our studies, we select two vortex beams with topological charges $m=5$ and $m=15$ that most clearly reflect properties of the sector perturbation. Computer simulation of typical vortex spectra is shown in Fig.3. We revealed a clear maximum of vortex mode intensity $\overline{C}^2_{m,n}$ for the initial topological charges $n=m=5$ and $m=n=15$ at the angle $\alpha = 45°$. Only a small part of the beam intensity is pumped over the neighboring vortex modes being distributed almost symmetrically relative to the maximum. However, as the angle $\alpha$ increases (or the adjacent angle $\beta = \pi - \alpha$ decreases), a breaking down of the symmetric intensity distribution among the vortex modes is observed. There is a distinct maximum in the spectrum range of negative topological charges $n < 0$ with the intensity maximum at $n = -5$ and $n = -15$. The mode spectral intensities at the region of positive $n > 0$ and negative $n < 0$ topological charges are nearly equalized at the angle $\beta = 2^0$. A characteristic feature of the intensity redistribution process between modes is the faster energy transfer between modes for vortex beams with lower topological charges (see Fig.3 for the angle $\beta = 6^o$ for $m = 5$ and $m = 15$). It is worth noting that the authors of Ref. [14], testing the uncertainty principle, also plotted the vortex spectra for a topologically neutral beam $m = 0$ and a beam with a small $m = 2$ at the angle $\alpha = 45^o$. The authors did not detect an intensity maximum in the negative region of topological charges. As we showed above, the appearance of the second spectral maximum is possible only at sufficiently small angles $\beta$ and relatively large topological charges $m$ of the initial vortex beam.

*The orbital angular momentum.* The OAM per photon of a complex perturbed beam is found as [22]

$$\ell_z(\alpha, m) = \sum_{n=-\infty}^{\infty} n\, \overline{C}^2_{m,n}(\alpha) \Big/ \sum_{n=-\infty}^{\infty} \overline{C}^2_{m,n}(\alpha). \qquad (5)$$

The mode amplitudes $\overline{C}^2_{m,n}$ in Eq.(5) are given by the normalized field $\Psi_m$ of the Laguerre-Gauss beams so that the squared amplitudes $C^2_{m,n}$ and $\overline{C}^2_{m,n}$ are obeyed a simple relation $\overline{C}^2_{m,n} = 2^{-|n|-2} |n|! C^2_{m,n}$. The OAM variations $\ell_z(\alpha, m)$ with increasing the aperture angle $\alpha$ (decreasing the adjacent angle $\beta$) is illustrated by Fig.4a for topological charges $m = 5, m = 10$ and $m = 15$ of the initial vortex beam. The orbital angular momentum remains almost equal to the initial OAM $\ell_z \approx m$ of the vortex beam in a wide range of angles $0 < \alpha < 7\pi / 8$ despite the rapid increase in the number of vortex states (see Fig.3). Then there is a sharp decrease of the OAM that is directly related to shaping the second spectral maximum in the negative region of topological charges in Fig.3. The angular momentum is reduced to zero already at an angle $\beta \approx 1^o$ that corresponds to the equalization of the beam intensity in the vicinity of two spectral maxima $\overline{C}^2_{m,n}$.

*Informational entropy (Shannon's entropy).* At first we note that the normalized squared amplitude $C^2_n \in (0,1)$ in the expansion (2) can be treated as a conditional probability $P(n/m)$ of encountering a vortex beam in the state $|n\rangle$ among $2N$ states, provided that the external perturbation $\alpha$ affected the vortex state $|m\rangle$ i.e. $P(n/m) = C^2_n(\alpha, m)$ (see e.g. [16,17]). Such an approach to counting a number of vortex states can be used in the Shannon formula [16]. Back in 1948, Claude Shannon proposed to take into account the measure of information loss (measure of uncertainty) that occurs when a perturbation acts on the information channel $m$ due to the redistribution of energy through other channels $n$ by the expression

$$S_I = -\sum_{n=0}^{N} P(n/m) \log_2 P(n/m) = -\sum_{n=0}^{N} C^2_n(\alpha, m) \log_2 C^2_n(\alpha, m). \qquad (6)$$

The entropy $S_I$ is a real positive magnitude that varies in a wide range $0 < S_I < \infty$ whereas the squared normalized amplitude changes in the range $1 > C^2_n(\alpha, m) > 0$. Fig.4b shows the dependence of the information entropy $S_I$ on the angle $\alpha$ for various topological charges of the initial vortex beam. Over a wide range of angles $0 < \alpha < 7\pi/8$, the entropy $S_I$ increases in the same manner for different values of topological charges $m$ although the OAM $\ell_z$ remains constant (see Fig.4a) that corresponds to the same features of changing a number of vortex states in Fig.3. However, the growth rate of the vortex number changes from slow to fast one at the angle $\alpha = \pi/2$ (that ccorresponds to cutting half the beam energy flow). At large angles $\alpha$ (small angles $\beta$), we observe a rapid divergence of the curves $S_I(\alpha)$ for different topological charges $m$, that is associated with the characteristic properties of the uncertainty principle between the angle and OAM. It is at this anglee range that the OAM in Fig. 4a rapidly decreases.

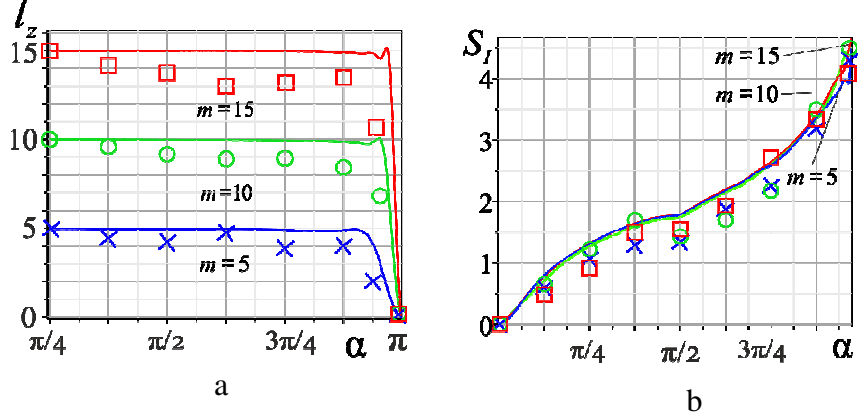

Fig.4 Computer simulation of (a) the OAM $\ell_z(\alpha,m)$ and (b) information entropy $S_I(\alpha,m)$ for the initial topological charges , $m=10$, $m=15$ (solid lines); crosslets $(\times)$, circlets $(\odot)$ and squares $(\square)$ are correspondent experimental data.

**2.** As we saw above, the vortex spectrum $C_{m,n}^2(\alpha)$ enables us to reconstruct both the OAM $\ell_z(\alpha,m)$ and informational entropy $S_I(\alpha,m)$ of a perturbed vortex beam. Therefore, we focused on experimental studies on measuring the spectrum of vortices. The main idea of the measurements was to transform single Laguerre-Gauss beams (LG) into single Hermite-Gauss beams (HG) via a cylindrical lens [22] as shown in Fig.1. A cylindrical lens can divide LG vortex beams with opposite signs of topological charges in such a way that LG beams with negative and positive $m$ correspond to HG beams with astigmatism axes perpendicular to each other. For the analysis of combined vortex beams, the intensity moments technique is used described in detail in Ref. [21]. For shaping vortex beams with a sector aperture, we refused to use sector holographic gratings [14] due to large measurement errors arising from astigmatic transformations. Instead, the metal sector aperture (D) was installed at the cylindrical lens plane (CL), while the appropriate vortex beam was shaped by a spatial light modulator (SLM). (see Fig.1).

The measurements of the vortex spectra $C_{m,n}^2(\alpha)$ were carried out for each sector perturbation $\alpha$ given by the aperture diaphragm (D). Inasmuch as a large array of optical vortices was involved in the measurement process, it was difficult to place in the same figure, both the experimental results and computer simulations we displayed the envelopes of the discrete spectrum $C_{m,n}^2(\alpha)$ in Fig.5. For a relatively small angles of the sector aperture, when a small number of vortices ($N \sim 11$, Fig.5a) are included in the process, we observe a slight mismatch of the discrete spectrum envelopes that corresponds to the measurement error $6\%$. The callouts in Fig. 5 illustrate the intensity distribution at the plane of the cylindrical lens and at the plane of its double focus. A nearly symmetric shape of the laser spot at the double focus plane in Fig. 5a (Cyl) points out to a small contribution of vortices with negative topological charges. However, a vertical orientation of the diffused laser spot in Fig. 5b (Cyl) indicates a significant contribution of negative charged vortices. For large sector angles $\alpha$, the uncertainty principle involves a large number of vortices in the process ($N \sim 50$ in Fig.5b), and the measurement error is reduced to 2%. We have noticed that the vortex spectrum shape is very sensitive to inaccuracies in the alignment of the sector aperture relative to the cylindrical lens. However, such sensitivity sharply decreases at large angles $\alpha$ of a sector perturbation. It is small inaccuracies in the alignment that lead to mismatches between the computer simulation and the OAM measurement results at small angles $\alpha$ in Fig. 4a.

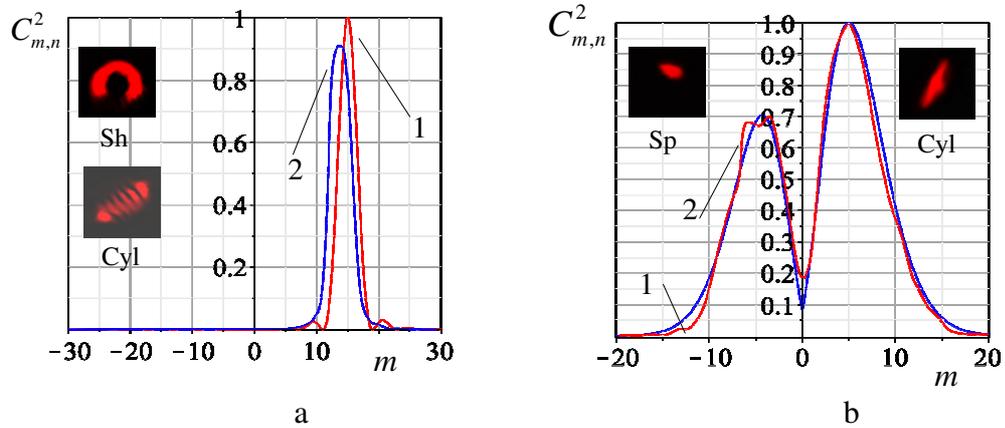

Fig.5 Envelopes of discrete vortex theoretical (1) and experimental (2) spectra of a perturbed vortex beam with an initial topological charge $m = 5$ at the sector angles (a) $\alpha = 45^o$ and (b) $\alpha = 173^o$. Callouts: snapshots of perturbed vortex beams at a cylindrical lens plane (Sp) and its double focus plane (Cyl)

Thus, we theoretically and experimentally showed that the sector perturbation, although it almost does not change the OAM of the vortex beam at small perturbation angles, the information entropy quickly grows that indicates an increase of the vortex number in the process.

## Acknowledgement

The authors thank Abramochkin E.G. for discussion and comments on the theoretical results.

## Abstract

We theoretically and experimentally investigated transformations of vortex beams subjected to sector perturbations in the form of hard-edged aperture. The transformations of the vortex spectra, the orbital angular momentum, and the informational entropy of the perturbed beam were studied. We found that relatively small angular sector perturbations have almost no effect on OAM, although the informational entropy is rapidly increasing due to the birth of new optical vortices caused by diffraction by diaphragm edges. At large perturbation angles, the uncertainty principle between the angle and OAM involves vortices, with both positive and negative topological charges, so that the OAM decreases to almost zero, and the entropy increases sharply.